\documentclass[a4paper,11pt]{article}
\begin{document}
\begin{titlepage}
\title{\bf
Itinerant Antiferromagnetism of Correlated Lattice Fermions
\thanks{Physica A267 (1999) 131-152}}
\author{
{\bf A.L.Kuzemsky\thanks{E-mail:kuzemsky@thsun1.jinr.ru;
http://thsun1.jinr.ru/~ kuzemsky}
}\\
Bogoliubov Laboratory of Theoretical Physics, \\
Joint Institute for Nuclear Research,\\
141980 Dubna, Moscow Region, Russia. }
\date{}
\maketitle
\begin{abstract}
The problem of finding of the ferromagnetic and antiferromagnetic
"symmetry broken" solutions of the correlated lattice fermion
models beyond the mean-field approximation  has been investigated.
The calculation of the quasiparticle excitation spectra with
damping for the single- and multi-orbital Hubbard model has been
performed in the framework of the equation- of-motion method for
two-time temperature Green's Functions within a non-perturbative
approach. A unified scheme for the construction of Generalized
Mean Fields (elastic scattering corrections) and self-energy
(inelastic scattering) in terms of the Dyson equation has been
generalized in order to include the presence of the "source
fields". The damping of quasiparticles, which reflects the
interaction of the single-particle and collective degrees of
freedom has been calculated.  The "symmetry broken" dynamical
solutions of the Hubbard model, which correspond to various types
of itinerant antiferromagnetism has been discussed. This approach
complements previous studies and clarifies the nature of the
concepts of itinerant antiferromagnetism and "spin-aligning field"
of correlated lattice fermions.
\end{abstract}
\end{titlepage}
\newpage
\section{Introduction}
The problem of the adequate description of  strongly correlated
lattice fermions has been studied intensively during the last
decade, especially in the context of Heavy Fermions and High-Tc
superconductivity~\cite{kuz1} -~\cite{kuz3}.  The behaviour and
the true nature of the electronic states and their quasiparticle
dynamics is of central importance to the understanding of the
magnetism in metals and the Mott-Hubbard metal-insulator
transition in oxides, the heavy fermions in rare-earths compounds
and the high-temperature superconductivity (HTSC) in cuprates.
Recently there has been considerable interest in identifying the
microscopic origin of these states~\cite{geb4}.  Antiferromagnetic
correlations may play an important role in the possible scenario
of normal and superconducting  behavior of these compounds. Some
of the experimental and theoretical results show that
antiferromagnetic spin fluctuations are really involved in the
problem.  This idea has stirred a great deal of discussion in
recent times~\cite{ruv5}. An appealing but phenomenological
picture of HTSC, known as the nearly antiferromagnetic Fermi
liquids (NAFL) approach, has been developed to explain many
anomalous properties of cuprates~\cite{pin6}. This approach
predicts the detailed phase diagram for cuprates~\cite{pin6} and
present arguments which suggest that the physical origin of the
pseudogap found in quasiparticle spectrum below the critical
temperature is the formation of a precursor to a
spin-density-wave-state. While the NAFL's scenario is appealing,
it has aparently not yet been derived from fully microscopic
consideration. The problem of the role of antiferromagnetic spin
fluctuations for HTSC has recently been the subject of many
papers( for a recent review see e.g.  Ref.~\cite{kampf7}).  These
investigations call for a better understanding of the nature of
solutions (especially magnetic) to the Hubbard and related
correlated models~\cite{kuz8} - \cite{kuz11}.  The microscopic
theory of the itinerant ferromagnetism and
antiferromagnetism~\cite{kem12},~\cite{joh13} of strongly
correlated fermions on a lattice at finite temperatures is one of
the important issues of recent efforts in the field~\cite{lieb14}
-~\cite{kampf17}. In some papers  the spin-density-wave (SDW)
spectrum was only used without careful and complete analysis of
the quasiparticle spectra of correlated lattice fermions.  The aim
of this paper is to investigate the intrinsic nature of the
"symmetry broken" ( ferro- and antiferromagnetic) solutions of the
Hubbard model at finite temperatures from the many-body point of
view. In the previous papers we set up the formalism and derived
the equations for the quasiparticle spectra with damping within
the single- and multi-orbital Hubbard model for the uniform
paramagnetic case. In this paper we apply the formalism to
consider the ferromagnetic and antiferromagnetic solutions. It is
the purpose of this paper to explore more fully the notion of
Generalized Mean Fields (GMF)~\cite{kuz10} which may arise in the
system of correlated lattice fermions to justify and understand
the "nature" of the local staggered mean-fields which fix the
antiferromagnetic ordering.  The present work brings together the
formulation of the itinerant antiferromagnetism of various papers.
For this aim we rederive the SDW spectra by the Irreducible
Green's Functions (IGF) method~\cite{kuz18} taking into account
the damping of quasiparticles.  This alternative derivation has a
close resemblance to that of the BCS theory of superconductivity
for transition metals~\cite{kuz19},~\cite{kuz20}  using the Nambu
representation (c.f.~\cite{raja21}).This aspect of the theory is
connected with the concept of broken symmetry, which is discussed
in detail for the present case. The advantage of the Green's
function method is the relative ease with which temperature
effects may be calculated.
\section{Itinerant Antiferromagnetism} The antiferromagnetic state is
characterized by a spatially changing component of magnetisation
which varies in such a way that the net magnetisation of the
system is zero. The concept of antiferromagnetism of localized
spins which is based on the Heisenberg model and the
two-sublattice Neel ground state is relatively well founded
contrary to the antiferromagnetism of delocalized or itinerant
electrons. The itinerant-electron picture is the alternative
conceptual picture for magnetism~\cite{her22}.\\ We now sketch the
main ideas of the concept of itinerant antiferromagnetism. The
simplified band model of an antiferromagnet  has been formulated
by Slater~\cite{slat23} within the single-particle Hartree-Fock
(H-F) approximation. In this approach he used the "exchange
repulsion" to keep electrons with parallel spins away from each
other and to lower the Coulomb interaction energy. Some authors
consider it as a prototype of the Hubbard model. However the
exchange repulsion was taken proportional to the number of
electrons with the same spins only and the energy gap between two
subbands was proportional to the difference of electrons with up
and down spins. In the antiferromagnetic many-body problem there
is an additional "symmetry broken" aspect.  For an
antiferromagnet, contrary to ferromagnet, the one-electron H-F
potential can violate the translational crystal symmetry. The
period of the antiferromagnetic spin structure $L$ is greater than
the lattice constant $a$. To introduce the two-sublattice picture
for itinerant model one should assume that $L=2a$ and that the
spins of outer electrons on neighboring atoms are antiparallel to
each other. In other words, the alternating (H-F) potential
$v_{i\sigma} = -\sigma v \exp(iQR_{i})$ where $Q =
(\pi/2,\pi/2,\pi/2)$ corresponds to a two-sublattice AFM
structure. To justify an antiferromagnetic ordering with
alternating up and down spin structure we must admit that in
effect two different charge distributions will arise concentrated
on atoms of sublattices A and B.  This  picture  accounts well for
quasi-localized magnetic behavior.\\ The earlier theories of
itinerant antiferromagnetism were proposed by des
Cloizeaux~\cite{declo24} and especially Overhauser~\cite{over25}
(in the context of the investigation of the ground state of
nuclear matter). Then Overhauser~\cite{over26} has applied this
approach for the explanation of the anomalous properties of dilute
$Cu-Mn$ alloys, has suggested an antiferromagnetic mechanism that
requires neither two-body interactions between paramagnetic solute
spins, nor a sublattice structure (c.f.~\cite{vig27}). Such a
mechanism may be recognized by considering a new type excited
state of the conduction electron gas. He invented the static SDW
which allow the total charge density of the gas to remain
spatially uniform.  Overhauser~\cite{over25} - \cite{ham29}
suggested that the H-F ground state of a three dimensional
electron gas is not necessarily a Slater determinant of plane
waves. Alternative sets of one-particle states can lead to a lower
ground-state energy.  Among these alternatives to the plane-wave
state are the SDW and CDW ground states for which the one-electron
Hamiltonians have the form \begin{equation} H = (p^{2}/2m) -
G(\sigma_{x} \cos Qz + \sigma_{y} \sin Qz) \end{equation} ( spiral
SDW; $Q = 2k_{F}z$ )\\ and\\ \begin{equation} H = (p^{2}/2m) -
2G\cos (Qr)
\end{equation} ( CDW; $Q = 2k_{F}z$ )\\ The periodic potentials in the
above expressions lead to a corresponding variation in the electronic
spin and charge densities, accompanied by a compensating variation of
the background. The effect of Coulomb interaction on the magnetic
properties of the electron gas in Overhauser's approach renders the
paramagnetic plane-wave state of the free-electron-gas model unstable
within the H-F approximation. The long-range components of the Coulomb
interaction are most important in creating this
instability~\cite{ham29}.  It was demonstrated~\cite{over28}  that a
nonuniform static SDW is lower in energy than the uniform (paramagnetic
state) in the Coulomb gas within the H-F approximation for certain
electron density.\\ The H-F is the simplest approximation but neglects
the important dynamical part. To include the dynamics one should take
into consideration the correlation effects. The role of correlation
corrections which  tend to suppress the SDW state as well as the role
of shielding and screening were not fully
clarified~\cite{bayl30}.  Overhauser remarked that SDW
ground states do not occur for $\delta$-function interactions, whatever
their strength.  This question was investigated further in
Ref.~\cite{berg31}.  An instability of the paramagnetic Hartree-Fock
state against a state with different orbitals for different spins  was
interpreted as a magnetic phase transition.\\ It is important to note
that in the Slater's and des Cloiseaux's models an electron moving in a
crystal does not change its spin.  In these models the main processes
are related with the pairing of electrons having the same spins, one
from each of the two sublattices. In the Overhauser's approach to
itinerant antiferromagnetism the combination of the electronic states
with different spins (with pairing of the opposite spins) is used to
describe the SDW state with period $Q$.  The first approach is
obviously valid only in the simple commensurate two-sublattice case and
the latter is applicable to the more general case of an incommensurate
spiral spin state. The general SDW state has the form
\begin{equation} \Psi_{p\sigma} = \chi_{p\sigma} \cos (\theta_{p}/2) +
\chi_{p+Q-\sigma} \sin (\theta_{p}/2) \end{equation} The average
spin for helical or spiral spin arrangement changes its direction
in the (x-y) plane.  For the spiral SDW states a spatial variation
of magnetisation corresponds to $\vec Q = (\frac{\pi}{a})(1,1)$.\\
The antiferromagnetic phase of
chromium~\cite{cade32},~\cite{faw33} and its alloys has been
satisfactorily explained in terms of the SDW within a two-band
model~\cite{fed34}.  It is essential to note that chromium becomes
antiferromagnetic in a unique manner. The antiferromagnetism is
established in a more subtle way from  the spins of the itinerant
electrons than  the magnetism of collective band electrons in
metals like iron and nickel. The essential feature of chromium
which makes possible the formation of the SDW is the existence of
"nested" portions of the Fermi surface~\cite{faw33}.  The
formation of bound electron-hole pairs takes place between
particles of opposite spins; the condensed state exhibits the
SDW.\\ The recent attempt to describe an antiferromagnetic
insulator at $T = 0$ using a one-electron approach was made in
Ref.~\cite{call35}.  To do this, the authors proposed to overcome
the inadequacies of standard local-spin-density theory by adding a
spin-dependent magnetic pseudopotential to Kohn-Sham equations.
\\For the Hubbard model~\cite{hub36} the qualitative phase diagram
was calculated by Penn~\cite{penn37}.  Unfortunately, although his
work gives a clear physical picture, it does not emphasize the
lattice character of the tight-binding or Wannier fermions as well
as the essence of the anomalous spin-flip averages. The Hubbard
model is a simplified but workable model for the correlated
lattice fermions and the applicability of the SDW Overhauser
concept to highly correlated tight binding electrons on a lattice
deserves a careful analysis within this model. In earlier
papers~\cite{adam38} - ~\cite{kris41} the single- and
multi-orbital Hubbard model has been investigated with respect to
antiferromagnetic solutions in the mean-field approximation
mainly. \section{ Hubbard Model} The Hubbard model has been widely
recognised as a workable model for a study of the correlated
itinerant electron systems. For the sake of completeness we shall
discuss the single-orbital and multi-orbital cases separately.
\subsection{MULTI-ORBITAL HUBBARD MODEL}
 To demonstrate
the advantage of our approach we shall consider the quasiparticle
spectrum of the lattice fermions for degenerate band model.  Let
us start with the second quantized form of the Hamiltonian taking
the set of the Wannier functions $[\phi_{\lambda}(r-R_{i})]$. Here
$\lambda$ is the band index ($\lambda$= 1,2,...5).
\begin{equation} H =
\sum_{ij\mu\nu\sigma}t^{\mu\nu}_{ij}a^{+}_{i\mu\sigma}a_{j\nu\sigma} +
\frac{1}{2}
\sum_{ij,mn}\sum_{\alpha\beta\gamma\delta\sigma\sigma'}
<i\alpha,j\beta|W|m\gamma,n\delta>
a^{+}_{i\alpha\sigma}a^{+}_{j\beta\sigma'}a_{m\gamma\sigma'}a_{n\delta\sigma}
\end{equation}
For a degenerate d-band the second
quantized form of the total Hamiltonian in the Wannier-function
representation  reduces to the following model Hamiltonian
\begin{equation} H = H_{1} + H_{2} + H_{3} \end{equation} The kinetic
energy operator is given by \begin{equation} H_{1} = \sum_{ij}
\sum_{\mu\nu\sigma} t^{\mu\nu}_{ij}a^{+}_{i\mu\sigma}a_{j\nu\sigma}
\end{equation} The term $H_{2}$ describes one-centre Coulomb
interactions \begin{eqnarray} H_{2} = \frac{1}{2} \sum_{i\mu\sigma}
U_{\mu\mu}n_{i\mu\sigma}n_{i\mu-\sigma} + \frac{1}{2}
\sum_{i\mu\nu}\sum_{\sigma\sigma'}
V_{\mu\nu}n_{i\mu\sigma}n_{i\nu\sigma'}(1 - \delta_{\mu\nu}) -\\
\nonumber
\frac{1}{2}\sum_{i\mu\nu\sigma}
I_{\mu\nu}n_{i\mu\sigma}n_{i\nu\sigma}(1 - \delta_{\mu\nu})
+
\frac{1}{2}\sum_{i\mu\nu\sigma}
I_{\mu\nu}a^{+}_{i\mu\sigma}a^{+}_{i\mu-\sigma}a_{i\nu-\sigma}a_{i\nu\sigma}
(1 - \delta_{\mu\nu}) -\\
\nonumber
\frac{1}{2} \sum_{i\mu\nu\sigma}
I_{\mu\nu}a^{+}_{i\mu\sigma}a_{i\mu-\sigma}a^{+}_{i\nu-\sigma}a_{i\nu\sigma}
(1 - \delta_{\mu\nu})
\end{eqnarray}
In addition to the intrasite intraorbital interaction $U_{\mu\mu}$ which
is the only interaction present in the single-orbital
Hubbard model, this term contains three more kinds of
interorbital interactions.\\ The last term  $H_{3}$
describes the direct intersite exchange interaction
\begin{equation}
H_{3} = -\frac{1}{2}
\sum_{ij\mu}
\sum_{\sigma\sigma'}
J^{\mu\mu}_{ij}
a^{+}_{i\mu\sigma}a_{i\mu-\sigma'}a^{+}_{j\mu\sigma'}a_{j\mu\sigma}
\end{equation}
The definition of various integrals in $H$ is obvious. It is reasonable
to assume that:
\begin{equation}
U_{\mu\mu} = U; \quad V_{\mu\nu} = V; \quad I_{\mu\nu} = I; \quad
J^{\mu\mu}_{ij} = J_{ij}.  \end{equation} This Hamiltonian differ
slightly from the analogous Hamiltonian of Ref.~\cite{oles40}
where the only intrasite interaction terms of the second-quantized
Hamiltonian of the $d$-band were taken into consideration.
\subsection{SINGLE-ORBITAL HUBBARD MODEL}
The model Hamiltonian which is
usually referred to as Hubbard Hamiltonian~\cite{hub36}
\begin{equation} H =
\sum_{ij\sigma}t_{ij}a^{+}_{i\sigma}a_{j\sigma} +
U/2\sum_{i\sigma}n_{i\sigma}n_{i-\sigma}
\end{equation}
includes the intraatomic Coulomb repulsion $U$ and the one-electron hopping
energy $t_{ij}$. The electron correlation forces electrons to localize in
the atomic orbitals, which are modelled here by the complete and orthogonal
set of the Wannier wave functions $[\phi({\vec r} -{\vec R_{j}})]$.
(The Wannier representation, which is a unitary transformation of the
Bloch representation is  an important background of the Hubbard model.
It is well known that in one-dimension the Wannier functions decrease
exponentially but less is known about two- and three-dimensional
cases.)
On the
other hand, the kinetic energy is reduced when electrons are delocalized.
The main difficulty of the right solution of the Hubbard
model is the necessity of taking into account both these effects
simultaneously. Thus, the Hamiltonian (10) is specified by two
parameters: $U$ and effective electron bandwidth $$\Delta =
(N^{-1}\sum_{ij}\vert t_{ij}\vert^{2})^{1/2}.$$ \\The important third
"player" is the Pauli principle, which has a long-range character,
contrary to the local Coulomb repulsion and nearest-neighbour hopping.
\\The band energy of Bloch electrons $\epsilon(\vec k)$ is defined as
follows $$t_{ij} = N^{-1}\sum_{\vec k}\epsilon(\vec k) \exp[i{\vec
k}({\vec R_{i}} -{\vec R_{j}})],$$ where  $N$ is the number of the
lattice sites. It is convenient to count the energy from the center of
gravity of the band, i.e.  $t_{ii} = \sum_{k}\epsilon(k) = 0$. The
effective electron bandwidth $\Delta$ and Coulomb intrasite integral
$U$ define completely the different regimes in 3 dimension depending on
parameter $\gamma = \Delta/U$. It is usually a rather difficult task to
find interpolating solution for the dynamical properties of the
Hubbard model.  We evidently have to to improve the early Hubbard's
theory taking account of the variety of possible regimes for the model
depending on electronic density, temperature and values of $\gamma$. It
was the purpose of the papers~\cite{kuz3}, ~\cite{kuz10} to find the
electronic quasiparticle spectra in a wide range of the temperature
and the parameters of the model  and to account explicitly for the
contribution of damping of the electronic states when calculating the
various characteristics of the model.  In the past years many
theoretical papers have been published , in which the approximative
dynamical solution of the models (5) and (10) has been investigated by
means of various advanced methods of many-body theory.  Despite the
considerable contributions to the development of the many-body theory
and to our better understanding of the physics of the correlated
electron systems, the fully consistent dynamical  analytical solution
of the Hubbard model is still lacking.  To solve this problem with a
reasonably accuracy and to describe correctly an interpolating solution
one need a more sophisticated approach than the usual procedures which
have been developed for description of the interacting electron-gas-
problem.  \section{Irreducible Green's Functions Method} Recent
theoretical investigations of strongly correlated electron systems have
brought forward a significant variety of  approaches.  To describe
from  first principles of the condensed matter theory and
statistical mechanics the physical properties of strongly correlated
systems we need to develop a systematic theory of quasiparticle
spectra.\\ In this paper we will use the approach which allows one to
describe completely the quasi-particle spectra with damping in a very
natural way.  This approach has been suggested to be essential for
various many-body systems and we believe that it bears the real physics
of strongly correlated electron systems~\cite{kuz10}, \cite{kuz18}.
The essence of our consideration of the dynamical properties of
many-body system with strong interaction is related closely with the
field theoretical approach and use the advantage of the Green's
functions language and the Dyson equation. It is possible to say that
our method tend to emphasize the fundamental and central role of the
Dyson equation for the single-particle dynamics of the many-body
systems at finite temperatures.\\ In this Section, we will discuss
briefly this novel nonperturbative approach for the description of the
many-body dynamics of strongly correlated systems.
A
number of other approaches has been proposed and  our approach is in
many respect additional and incorporates the logic of development of
the many-body techniques.  The considerable progress in studying the
spectra of elementary excitations and thermodynamic properties of
many-body systems has been for most part due to the development of the
temperature dependent Green's Functions methods. We have developed a
helpful reformulation of the two-time GFs method which is especially
adjusted~\cite{kuz3} for the correlated fermion systems on a lattice.
The very important concept of the whole method are the {\bf Generalized
Mean Fields}. These GMFs have a complicated structure for the strongly
correlated case and do not reduce to the functional of the mean
densities of the electrons, when we calculate excitations spectra at
finite temperatures.  To clarify the foregoing, let us consider the
retarded GF of the form \begin{equation} G^{r} = <<A(t), B(t')>> =
-i\theta(t - t')<[A(t)B(t')]_{\eta}>, \eta = \pm 1.  \end{equation} As
an introduction of the concept of IGFs let us describe the main ideas
of this approach in a symbolic form. To calculate the retarded GF $G(t
- t')$ let us write down the equation of motion for it:
\begin{equation} \omega G(\omega) = <[A, A^{+}]_{\eta}> + <<[A,
H]_{-}\mid A^{+}>>_{\omega}.  \end{equation} The essence of the method
is as follows~\cite{kuz18}. It is based on the notion of the {\it
``IRREDUCIBLE"} parts of GFs (or the irreducible parts of the
operators, out of which the GF is constructed) in term of which it is
possible, without recourse to a truncation of the hierarchy of equations
for the GFs, to write down the exact Dyson equation and to obtain an
exact analytical representation for the self-energy operator. By definition
we introduce the irreducible part {\bf (ir)} of the GF
\begin{equation}
^{ir}<<[A, H]_{-}\vert A^{+}>> = <<[A, H]_{-} - zA\vert A^{+}>>.
\end{equation}
The unknown constant z is defined by the condition (or constraint)
\begin{equation}
<[[A, H]^{ir}_{-}, A^{+}]_{\eta}> = 0
\end{equation}
From the condition (14) one can find:
\begin{equation}
z = \frac{<[[A, H]_{-}, A^{+}]_{\eta}>}{<[A, A^{+}]_{\eta}>} =
 \frac{M_{1}}{M_{0}}
\end{equation}
Here $M_{0}$ and $M_{1}$ are the zeroth and first order moments of the
spectral density. Therefore, irreducible GF  are defined so that they
cannot be reduced to the lower-order ones by any kind of decoupling. It
is worthy to note that the irreducible correlation functions are well
known in statistical mechanics. In the diagrammatic approach the
irreducible vertices are defined as the graphs that do not contain
inner parts connected by the $G^{0}$-line. With the aid of the definition
(13) these concepts are translated into the language of retarded and
advanced GFs. This procedure extract all relevant (for the problem under
consideration) mean field contributions and puts them into the generalized
mean-field GF, which here are defined as
\begin{equation}
G^{0}(\omega) = \frac{<[A, A^{+}]_{\eta}>}{(\omega - z)}.
\end{equation}
To calculate the IGF $ ^{ir}<<[A, H]_{-}(t), A^{+}(t')>>$ in (12), we
have to write the equation of motion after differentiation with respect
to the second time variable $t'$. The condition (14) removes the
inhomogeneous term from this equation and is a very crucial point of
the whole approach. If one introduces an irreducible part for the
right-hand side operator as discussed above for the ``left" operator,
the equation of motion (12) can be exactly rewritten in the following
form \begin{equation} G = G^{0} + G^{0}PG^{0}.  \end{equation} The
scattering operator $P$ is given by \begin{equation} P =
(M_{0})^{-1}\quad ^{ir}<<[A, H]_{-}\vert[A^{+}, H]_{-}>>^{ir}
(M_{0})^{-1}.
\end{equation}
The structure of the equation (17) enables us to determine the
self-energy operator $M$, in complete analogy with the diagram
technique \begin{equation} P = M + MG^{0}P.  \end{equation} From the
definition (19) it follows that  the self-energy
operator $M$ is defined as a proper (in diagrammatic language
``connected") part of the scattering operator $M = (P)^{p}$. As a
result, we obtain the exact Dyson equation for the thermodynamic
two-time Green's Functions:  \begin{equation} G = G^{0} + G^{0}MG,
\end{equation}
which has a well known formal solution of the form
\begin{equation}
G = [ (G^{0})^{-1} - M ]^{-1}
\end{equation}
Thus, by introducing irreducible parts of GF (or the irreducible parts of
the operators, out of which the GF is constructed) the equation of motion
(12) for the GF can be exactly (but using constraint (14)) transformed
into Dyson equation for the two-time thermal GF. This is very
remarkable result, which deserves  underlining, because of the
traditional form of the GF method did not include  this point.
The projection operator technique
has essentially the same philosophy, but with using the constraint (14)
in our approach we emphasize the fundamental and central role of the
Dyson equation for the calculation of the single-particle properties of
the many-body systems. It is important to note, that for the retarded
and advanced GFs the notion of the proper part is symbolic in
nature~\cite{kuz18}.
However, because of the identical form of the equations for the GFs for all
three types (advanced, retarded and causal), we can convert in each stage
of calculations to causal GFs and, thereby, confirm the substantiated nature
of definition (19)! We therefore should speak of an analogue of the
Dyson equation. Hereafter we will drop this stipulation, since it will
not cause any misunderstanding. It should be emphasized that the scheme
presented above give just an general idea of the IGF method. The
specific method of introducing IGFs depends on the form of operator
$A$, the type of the Hamiltonian and the conditions of the problem. The
general philosophy of the IGF method lies in the separation and
identification of elastic scattering effects and inelastic ones. This
last point is quite often underestimated and both effects are mixed.
However, as far as the right definition of quasiparticle damping is
concerned, the separation of elastic and inelastic scattering processes
is believed to be crucially important for the many-body systems with
complicated spectra and strong interaction. Recently it was emphasized
especially that the anomalous damping of electrons (or holes)
distinguishes cuprate superconductors from ordinary metals.  From a
technical point of view the elastic (GMF) renormalizations can exhibit
a quite non-trivial structure. To obtain this structure correctly, one
must construct the full GF from the complete algebra of the relevant
operators and develop a special projection procedure for higher-order
GF in accordance with a given algebra.  It is necessary to emphasize
that that there is an intimate connection between the adequate
introduction of mean fields and internal symmetries of the
Hamiltonian.
\section{Symmetry Broken Solutions} In many-body
interacting systems, the symmetry is important in classifying of
the different phases and in understanding of the phase transitions
between them~\cite{bog42} - ~\cite{tas48}. According to
Bogolubov~\cite{bog42}( c.f.~\cite{mat47} ) in each condensed phase, in
addition to the normal process, there is an anomalous process (or
processes) which can take place because of the long-range internal
field, with a corresponding propagator. The anomalous propagators for
interacting many-fermion system corresponding to the ferromagnetic (FM)
and antiferromagnetic (AFM) long-range ordering are given by
\begin{eqnarray} FM: G_{fm} \sim <<a_{k\sigma};a^{+}_{k-\sigma}>> \\
\nonumber AFM: G_{afm} \sim <<a_{k+Q\sigma};a^{+}_{k+Q'\sigma'}>>
\end{eqnarray} In the SDW case, a particle picks up momentum $Q - Q'$
from scattering against the periodic structure of the spiral
(nonuniform) internal field, and has its spin changed from $\sigma$ to
$\sigma'$ by the spin-aligning character of the internal field.  The
Long-Range-Order (LRO) parameters are:  \begin{eqnarray} FM: m =
1/N\sum_{k\sigma} <a^{+}_{k\sigma}a_{k-\sigma}>\\ \nonumber AFM: M_{Q}
= \sum_{k\sigma} <a^{+}_{k\sigma}a_{k+Q-\sigma}> \end{eqnarray} It is
important to note that the long-range order parameters are
functions of the internal field, which is itself a function of the
order parameter. There is a more mathematical way of formulating this
assertion. According to the paper~\cite{lieb14}, the notion "symmetry
breaking" means that the state fails to have the symmetry that the
Hamiltonian has.
True broken symmetry can arise only if there are infinitesimal "source
fields" present.  Indeed, for the rotationally and translationally
invariant Hamiltonian the suitable source terms should be added:
\begin{eqnarray} FM:  \varepsilon\mu_{B}
H_{x}\sum_{k\sigma}a^{+}_{k\sigma}a_{k-\sigma}\\ \nonumber AFM:
\varepsilon \mu_{B} H \sum_{kQ} a^{+}_{k\sigma}a_{k+Q-\sigma}
\end{eqnarray} where $\varepsilon \rightarrow 0$ is to be taken at
the end of calculations.\\ Broken symmetry solutions of the
Overhauser type (3) imply that the vector $Q$ is a measure of the
inhomogeneity or breaking of translational symmetry. It is
interesting to note the remark of
paper~\cite{tas46}(c.f.~\cite{tas48}) about antiferromagnetism,
for which "a staggered magnetic field plays the role of
symmetry-breaking field. No mechanism can generate a real
staggered magnetic field in an antiferromagnetic material".  The
Hubbard model is a very interesting tool for the analyzing  this
concept~\cite{kris41} - \cite{oza51}.\\ Penn~\cite{penn37} shown
that antiferromagnetic state and more complicated states (e.g.
ferrimagnetic) can be made eigenfunctions of the self-consistent
field equations within an "extended"  mean-field approach,
assuming that the "anomalous" averages
$<a^{+}_{i\sigma}a_{i-\sigma}>$ determine the behavior of the
system on the same footing as the "normal" density of
quasiparticles $<a^{+}_{i\sigma}a_{i\sigma}>$.  It is clear,
however, that these "spin-flip" terms break the rotational
symmetry of the Hubbard Hamiltonian. For the single- band Hubbard
Hamiltonian the averaging $<a^{+}_{i-\sigma}a_{i, \sigma}> = 0$
because of the rotational symmetry of the Hubbard model.  The
inclusion of the "anomalous" averages lead to the unresricted H-F
approximation. The rigorous definition of the unrestricted
Hartree-Fock approximation (UHFA) has been done recently in
Ref.~\cite{lieb14}.  This approximation has been applied also for
the single-band Hubbard model (10) for the calculation of the
density of states. The following definition of UHFA has been used:
\begin{equation} n_{i-\sigma}a_{i\sigma} =
<n_{i-\sigma}>a_{i\sigma} -
<a^{+}_{i-\sigma}a_{i\sigma}>a_{i-\sigma} \end{equation} Thus, in
addition to the standard H-F term, the new, the so-called
``spin-flip" terms, are retained. This example clearly show that
the nature of the mean-fields follows from the essentials of the
problem and should be defined in a proper way.  So, one needs a
properly defined effective Hamiltonian $H_{\rm eff}$.  We shall
analyze below in detail the proper definition of the irreducible
GFs which include the ``spin-flip" terms. For the single-orbital
Hubbard model this definition should be modified in the following
way:  \begin{eqnarray}
^{ir}<<a_{k+p\sigma}a^{+}_{p+q-\sigma}a_{q-\sigma} \vert
a^{+}_{k\sigma}>>_ {\omega} =
<<a_{k+p\sigma}a^{+}_{p+q-\sigma}a_{q-\sigma}\vert
a^{+}_{k\sigma}>>_{\omega} - \nonumber\\ \delta_{p,
0}<n_{q-\sigma}>G_{k\sigma} - <a_{k+p\sigma}a^{+}_{p+q-\sigma}>
<<a_{q-\sigma} \vert a^{+}_{k\sigma}>>_{\omega} \end{eqnarray}
From this definition it follows that such way of introduction of
the IGF broadens the initial algebra of the operators and the
initial set of the GFs.  This means that ``actual" algebra of the
operators must include the spin-flip terms at the beginning,
namely:  $(a_{i\sigma}$, $a^{+}_{i\sigma}$, $n_{i\sigma}$,
$a^{+}_{i\sigma}a_{i-\sigma})$. The corresponding initial GF will
have the form
$$\pmatrix{
<<a_{i\sigma}\vert a^{+}_{j\sigma}>> & <<a_{i\sigma}\vert
a^{+}_{j-\sigma}>> \cr <<a_{i-\sigma}\vert a^{+}_{j\sigma}>> &
<<a_{i-\sigma}\vert a^{+}_{j-\sigma}>> \cr}$$ With this definition
we introduce the so-called anomalous (off-diagonal) GFs which fix
the relevant vacuum and select the proper symmetry broken
solutions. In fact, this approximation has been investigated
earlier by Kishore and Joshi~\cite{kis50}. They clearly pointed
out that they assumed that the system is magnetized in the $x$
direction instead of the conventional $z$ axis.  The detailed
investigation and classification of the magnetic and non-magnetic
symmetry broken solutions of the three-band extended Hubbard model
for $CuO_{2}$ planes of high-$T_c$ superconductors was made in
Ref.~\cite{oza51} within the mean-field approximation.
\section{Dynamical Properties} In many-body interacting systems the
quasiparticle dynamics can be quite non-trivial.  Here the problem
of the adequate description of the many-body dynamics of the
multi-orbital Hubbard model  will be discussed in the framework of
the equation-of-motion approach for two-time thermodynamic Green's
Functions. Our main motivation was the intention to formulate a
consistent theory of dynamical properties of the Hubbard model
taking into account the symmetry broken (magnetic) solutions.\\
This formulation gives us an opportunity to emphasize some
important issues about the relevant dynamical solutions of the
strongly correlated  models of fermions on a lattice and to
formulate in a more sharp form the ideas of the method of the
Irreducible Green's Functions (IGF)~\cite{kuz18}.  This IGF method
allows one to describe the quasiparticle spectra with damping of
the strongly correlated electron systems in a very general and
natural way and to construct the relevant dynamical solution in a
self-consistent way on the level of the Dyson equation without
decoupling the chain of the equation of motion for the GFs.\\ The
interplay and the competition of the kinetic energy and potential
energy affects substantially the electronic spectrum.  The
renormalized electron energies are temperature dependent and the
electronic states have  finite life times. These effects  are most
suitably accounted for by the Green's functions method.  We shall
use the (IGF) method of Section 4.  To give a more instructive
discussion let us consider the single- particle GF of lattice
fermions, which is defined as \begin{eqnarray}
G^{\mu\nu}_{\sigma\sigma'}(ij;t - t') =
<<a_{i\mu\sigma}(t),a^{+}_{j\nu\sigma'}(t')>> = -i\theta(t -
t')<[a_{i\mu\sigma}(t),a^{+}_{j\nu\sigma'}(t')]_{+}> = \nonumber\\
1/2\pi \int_{-\infty}^{+\infty} d\omega \exp(-i\omega t)
G^{\mu\nu}_{\sigma\sigma'}(ij;\omega) \end{eqnarray} Actually,
this GF is a matrix (10x10) in the joint tensor product vector
space of spin and orbital momentum. The diagonal elements of this
matrix GF are normal propagators, while the off-diagonal elements
are anomalous.  The equation of motion for the Fourier transform
of the GF has the form \begin{eqnarray} \sum_{m\alpha}
A^{\mu\alpha}(im)G^{\alpha\nu}_{\sigma\sigma'}(mj;\omega) =
\delta_{ij} \delta_{\mu\nu} \delta_{\sigma\sigma'} +
\sum_{m\alpha}[ B^{\mu\alpha}_{1}(im)
<<a_{m\mu\sigma}n_{m\alpha\sigma}|a^{+}_{j\nu\sigma'}>>\\
\nonumber +
B^{\mu\alpha}_{2}(im)
<<a_{m\mu\sigma}n_{m\alpha-\sigma}|a^{+}_{j\nu\sigma'}>> + \\
\nonumber
B^{\mu\alpha}_{3}(im)
(<<a_{i\mu\sigma}n_{m\mu\sigma}|a^{+}_{j\nu\sigma'}>> +
<<a_{i\mu-\sigma}a^{+}_{m\mu-\sigma}a_{m\mu\sigma}|a^{+}_{j\nu\sigma'}>>)]
\end{eqnarray}
Here we have introduced the notations
\begin{eqnarray}
A^{\mu\alpha}(im) = \omega\delta_{mi}\delta_{\mu\alpha} -
t^{\mu\alpha}_{im}; \quad B^{\mu\alpha}_{1}(im) = (V -I) \delta_{im}(1
- \delta_{\mu\alpha});\\
\nonumber B^{\mu\alpha}_{2} = [U\delta_{\mu\alpha} + V(1 -
\delta_{\mu\alpha}]\delta_{im};  \quad B^{\mu\alpha}_{3}(im) =
J_{im}(1 - \delta_{im})\delta_{\mu\alpha} \end{eqnarray} Let us
introduce, by definition, an "irreducible" GF in the following way
\begin{eqnarray} (^{ir}<<a_{i\beta\sigma}
a^{+}_{m\alpha\sigma_{1}} a_{m\alpha\sigma_{1}}|
a^{+}_{j\nu\sigma'}>>) = <<a_{i\beta\sigma}
a^{+}_{m\alpha\sigma_{1}} a_{m\alpha\sigma_{1}}|
a^{+}_{j\nu\sigma'}>>
\\
\nonumber
-<n_{m\alpha\sigma_{1}}>\delta_{mi}
<<a_{i\beta\sigma}|a^{+}_{j\nu\sigma'}>> -
<a_{i\beta\sigma}a^{+}_{m\alpha\sigma_{1}}>
<<a_{m\alpha\sigma_{1}}|a^{+}_{j\nu\sigma'}>>
\end{eqnarray}
According to (14), the following constraint should be valid
\begin{equation}
<[(a_{i\beta\sigma}n_{m\alpha\sigma_{1}})^{(ir)},
a^{+}_{j\nu\sigma'}]_{+}> = 0
\end{equation}
Substituting (30) in (28) we obtain the following equation of motion
in the matrix (in spin space) form \begin{eqnarray}
\sum_{m\alpha}F^{\mu\alpha}(im)G^{\alpha\nu}(mj;\omega) =1 +
\sum_{m\alpha}[L^{\mu\alpha}_{1}(il)D^{\mu\alpha\,\nu}_{1}(mj) +\\
\nonumber
L^{\mu\alpha}_{2}(im)D^{\mu\alpha,\nu}_{2}(mj) + L^{\mu\alpha}_{3}(im)
(R^{\alpha\nu}_{1}(im,j) + R^{\alpha\nu}_{2}(im,j))]
\end{eqnarray}
where
\begin{eqnarray}
F^{\mu\alpha}(im) = \pmatrix{
E^{\mu\alpha}_{11}(im)&E^{\mu\alpha}_{12}(im)\cr
E^{\mu\alpha}_{21}(im)&E^{\mu\alpha}_{22}(im)\cr};
\quad 1 = \pmatrix{
1&0\cr
0&1\cr}\delta_{\mu\nu}\delta_{ij}\\
\nonumber
L^{\mu\alpha}_{1}(im) = \pmatrix{
B^{\mu\alpha}_{1}(im)&B^{\mu\alpha}_{2}\cr
0&0\cr}; \quad L^{\mu\alpha}_{2}(im) = \pmatrix{
0&0\cr
B^{\mu\alpha}_{1}(im)&B^{\mu\alpha}_{2}(il)\cr};\\
\nonumber
L^{\mu\alpha}_{3}(im) = \pmatrix{
B^{\mu\alpha}_{3}(im)&0\cr
0&B^{\mu\alpha}_{3}(im)\cr}
\end{eqnarray}
and
\begin{eqnarray}
E^{\mu\alpha}_{11}(im) = A^{\mu\alpha}(im) - B^{\mu\alpha}_{1}(im)
<a_{m\mu\uparrow}a^{+}_{m\alpha\uparrow}> - \\
\nonumber
\sum_{\beta}(B^{\mu\beta}_{1}(im)<n_{m\alpha\uparrow}>\delta_{\mu\beta}
- \\
\nonumber
-B^{\mu\beta}_{2}(im)<n_{i\alpha\downarrow}>\delta_{\mu\beta}) -\\
\nonumber
B^{\mu\alpha}_{3}(im)(<a_{i\alpha\uparrow}a^{+}_{m\alpha\uparrow}> +
<a_{i\alpha\downarrow}a^{+}_{m\alpha\downarrow}>) -
\sum_{l}B^{\mu\alpha}_{3}(ml)<n_{l\alpha\uparrow}>;\\
E^{\mu\alpha}_{12} =
-B^{\mu\alpha}_{2}<a_{m\mu\uparrow}a^{+}_{m\alpha\downarrow}> -
\sum_{l}
B^{\mu\alpha}_{3}<a^{+}_{l\alpha\downarrow}a_{l\alpha\uparrow}>\delta_{im}
\end{eqnarray}
and similar expressions for $E_{21}$ and $E_{22}$ with reversed spin
indices. The higher-order GF have the form
\begin{eqnarray}
\nonumber
D_{1} = \pmatrix{
(^{(ir)}<<a_{m\mu\uparrow}n_{m\alpha\uparrow}|a^{+}_{j\nu\uparrow}>>)&
(^{(ir)}<<a_{m\mu\uparrow}n_{m\mu\uparrow}|a^{+}_{j\nu\downarrow}>>)\cr
(^{(ir)}<<a_{m\mu\uparrow}n_{m\alpha\downarrow}|a^{+}_{j\nu\uparrow}>>)&
(^{(ir)}<<a_{m\mu\uparrow}n_{m\alpha\downarrow}|
a^{+}_{j\nu\downarrow}>>)\cr}\\
\nonumber
D_{2} = \pmatrix{
(^{(ir)}<<a_{i\alpha\uparrow}n_{m\alpha\uparrow}|a^{+}_{j\nu\uparrow}>>)&
(^{(ir)}<<a_{i\alpha\uparrow}n_{m\alpha\uparrow}|a^{+}_{j\nu\downarrow}>>)\cr
(^{(ir)}<<a_{i\alpha\downarrow}n_{m\alpha\downarrow}|a^{+}_{j\nu\uparrow}>>)&
(^{(ir)}<<a_{i\alpha\downarrow}n_{m\alpha\downarrow}|
a^{+}_{j\nu\downarrow}>>)\cr}\\
\end{eqnarray}
and $R$ has the following structure
$$
\begin{array}{l}
R = \\
\pmatrix{
(^{(ir)}<<a_{m\mu\uparrow}a^{+}_{m\alpha\uparrow}
a_{m\alpha\uparrow}|a^{+}_{n\nu\uparrow} >>)&
(^{(ir)}<<a_{m\mu\uparrow}a^{+}_{m\alpha\uparrow}
a_{m\alpha\uparrow}|a^{+}_{n\nu\downarrow}>>) \cr
(^{(ir)}<<a_{m\mu\uparrow}a^{+}_{m\alpha\downarrow}
a_{m\alpha\downarrow}|a^{+}_{n\nu\uparrow}>>)&
(^{(ir)}<<a_{m\mu\uparrow}a^{+}_{m\alpha\downarrow}
a_{m\alpha\downarrow}|
a^{+}_{n\nu\downarrow}>>) \cr}
\end{array}
$$
To calculate the higher-order GF $D_{1}$, $D_{2}$, $R_{1}$ and $R_{2}$,
we will differentiate the r.h.s. of it with respect to
the second-time variable (t'). Combining both (the
first- and second-time differentiated) equations of
motion we get the "exact"( no approximation have been made till now)
"scattering" equation
\begin{equation}
G^{\mu\nu}(ij;\omega) = G^{\mu\nu}_{0}(ij;\omega) +
\sum_{mn\alpha\beta}G^{\mu\alpha}_{0}(im;\omega)
P^{\alpha\beta}(mn;\omega) G^{\beta\nu}_{0}(nj;\omega)
\end{equation}
Here we have introduced the generalized mean-field (GMF) GF $G_{0}$
according to the following definition
\begin{equation}
\sum_{m\alpha}F^{\mu\alpha}(im)G^{\alpha\nu}_{0}(mj;\omega) =
 \delta_{ij}
\delta_{\mu\nu}
\end{equation}
The scattering operator $P$ has the form
\begin{equation}
P^{\mu\alpha}(mn;\omega) = \pmatrix{
P^{\mu\alpha}_{11}(mn;\omega)&P^{\mu\alpha}_{12}(mn;\omega) \cr
P^{\mu\alpha}_{21}(mn;\omega)&P^{\mu\alpha}_{22}(mn;\omega)\cr};
\end{equation}
Let us write down explicitly the first matrix element
\begin{eqnarray}
P^{\alpha\beta}_{11}(mn;\omega) =
\sum_{ij\mu\nu}[
B^{\mu\alpha}_{1} (im) (^{(ir)}
<<a_{m\mu\uparrow}n_{m\alpha\uparrow}
|a^{+}_{n\nu\uparrow}n_{n\beta\uparrow}>>^{(ir)})B^{\mu\beta}_{1}(nj)
+\\
\nonumber
B^{\mu\alpha}_{1} (im) (^{(ir)}
<<a_{m\mu\uparrow}n_{m\alpha\uparrow}
|a^{+}_{n\nu\uparrow}n_{n\beta\downarrow}>>^{(ir)})B^{\beta\nu}_{2}(nj)
+\\
\nonumber
B^{\mu\alpha}_{2} (im) (^{(ir)}
<<a_{m\mu\uparrow}n_{m\alpha\downarrow}
|a^{+}_{n\nu\uparrow}n_{n\beta\uparrow}>>^{(ir)})B^{\beta\nu}_{1}(nj)
+\\
\nonumber
B^{\mu\alpha}_{2} (im) (^{(ir)}
<<a_{m\mu\uparrow}n_{m\alpha\downarrow}
|a^{+}_{n\nu\uparrow}n_{n\beta\downarrow}>>^{(ir)})B^{\beta\nu}_{2}(nj)]
\\
\nonumber
\end{eqnarray}
Here we presented for brevity the
explicit expression for a part of Hamiltonian (5) only without the last
term.
Using (17) - (19) we find the Dyson equation in the Wannier basis
\begin{equation}
G^{\mu\nu}(ij;\omega) = G^{\mu\nu}_{0}(ij;\omega) +
\sum_{mn\alpha\beta}
G^{\mu\alpha}_{0}(im;\omega)M^{\alpha\beta}(mn;\omega)G^{\beta\nu}(nj;\omega)
\end{equation}
The equation (41) is the central result of the present treatment.
\section{Quasiparticle Formulation}
Let us first consider how to describe our system in terms of
quasiparticles. For a translationally invariant system, to
describe the low-lying excitations in terms of quasiparticles one
has to make a Fourier transformation
\begin{eqnarray}
G^{\mu\nu}(ij;\omega) = N^{-1}\sum_{k} \exp [ik(R_{i} -
R_{j})]G^{\mu\nu}(k;\omega)\\
\nonumber
M^{\mu\nu}(ij;\omega) = N^{-1}\sum_{k} \exp [ik(R_{i} -
R_{j})]M^{\mu\nu}(k;\omega)\\
\nonumber
t^{\mu\mu}_{ij} = N^{-1}\sum_{k} \exp [ik(R_{i} -
R_{j})]\epsilon_{\mu}(k)
\end{eqnarray}
The Dyson equation (41) in the Bloch vector space is given by
\begin{equation}
G^{\mu\nu}(k;\omega) = G^{\mu\nu}_{0}(k;\omega) +
\sum_{\alpha\beta}
G^{\mu\alpha}_{0}(k;\omega)M^{\alpha\beta}(k;\omega)G^{\beta\nu}(k;\omega)
\end{equation}
The renormalized energies in the mean field approximations are the
solutions of the equation
\begin{equation}
\sum_{\alpha}F^{\mu\alpha}(k)G^{\alpha\nu}_{0}(k;\omega) = 1
\delta_{\mu\nu}
\end{equation}
Using (44) we find
\begin{eqnarray}
E^{\alpha\nu}_{11}(k) = [\omega -
\epsilon_{\alpha}(k)]\delta_{\alpha\nu} - (1 - \delta_{\alpha\nu})(V -
I) K^{\alpha\nu}_{\uparrow\uparrow} -\\
\nonumber
\sum_{\mu}[(1 -
\delta_{\alpha\mu})\delta_{\alpha\nu}(V - I)N^{\mu}_{\uparrow} +
(U\delta_{\alpha\mu} + V(1 -
\delta_{\alpha\mu}))\delta_{\alpha\nu}N^{\mu}_{\downarrow}];\\
E^{\alpha\nu}_{12}(k) = [U\delta_\alpha\nu + V(1 - \delta_{\alpha\nu})]
K^{\alpha\nu}_{\uparrow\downarrow};\\
N^{\alpha}_{\sigma} =
N^{-1}\sum_{p}<a^{+}_{p\alpha\sigma}a_{p\alpha\sigma}>;\\
K^{\alpha\beta}_{\sigma_1\sigma_2} =
N^{-1}\sum_{p}<a_{p\alpha\sigma_1}a^{+}_{p\beta\sigma_2}>
\end{eqnarray}
For the degenerate Hubbard model ($V = I = J = 0$) we get
\begin{equation}
E^{\alpha\nu}_{11}(k) = [\omega -
\epsilon_{\alpha}(k) - UN^{\alpha}_{\downarrow}]\delta_{\alpha\nu}
\end{equation}
The spectrum of electronic low-lying excitations without damping
follows from the poles of the single-particle  mean-field GF
\begin{equation} \pmatrix{ \hat E_{11}&\hat E_{12}\cr \hat
E_{21}&\hat E_{22}\cr} \pmatrix{ \hat G_{011}&\hat G_{012}\cr \hat
G_{021}&\hat G_{022}\cr} = \pmatrix{ 1&0\cr 0&1\cr} \end{equation}
Here $\hat G_{0}$ denotes a matrix in the space of band indices.
If we put the spin-flip contributions equal to zero, i.e.
$$\hat E_{12} = \hat E_{21} = 0$$
then the quasiparticle spectra are given by
$$det|\hat E_{11}| = 0; \quad det|\hat E_{22}| = 0$$
For the  multiorbital Hubbard model (5) we find
\begin{equation}
G_{011}^{\alpha}(\omega) =[
\omega -\epsilon_{\alpha}(k) - UN^{\alpha}_{\downarrow}  -
V\sum_{\nu}(1 - \delta_{\alpha\nu})(N^{\nu}_{\downarrow} +
N^{\nu}_{\uparrow}) + I\sum_{\nu}(1 -
\delta_{\alpha\nu})N^{\nu}_{\uparrow}]^{-1}
\end{equation}
Finally we turn to the calculation of the damping.
To find the damping of the electronic states in the general case,
one needs to find the matrix elements of the self-energy in (43). Thus
we have \begin{equation} \pmatrix{ \hat G_{11}&\hat G_{12}\cr \hat
G_{21}&\hat G_{22}\cr} =\left[ \pmatrix{ \hat G_{011}&\hat G_{012}\cr
\hat G_{021}&\hat G_{022}\cr}^{-1} - \pmatrix{\hat M_{11}&\hat M_{12}\cr
\hat M_{21}&\hat M_{22}\cr}\right]^{-1} \end{equation}
From this matrix equation we have
\begin{eqnarray}
\hat G_{11} = (\hat G^{-1}_{011} - \hat \Sigma_{11})^{-1}; \quad
\hat G_{21} = (\hat G^{-1}_{021} - \hat \Sigma_{21})^{-1};\\
\nonumber
\hat G_{12} = (\hat G^{-1}_{012} - \hat \Sigma_{12})^{-1}; \quad
\hat G_{22} = (\hat G^{-1}_{022} - \hat \Sigma_{22})^{-1};
\end{eqnarray}
where the true self-energy has the form
\begin{equation}
\hat \Sigma_{11} = \hat M_{11} - \hat E_{12} \hat E^{-1}_{22}\hat M_{21}
+
\left [\hat M_{12} \hat E^{-1}_{22} + (\hat M_{12} - \hat E_{12}) \hat
E^{-1}_{22} \hat M_{22}(\hat E_{22} - \hat M_{22})^{-1} \right ] (\hat
M_{21} - \hat E_{21}) \end{equation}
The elements of the mass operator matrix $\hat M$ are proportional to
the higher-order GF of the following form
$$
(^{(ir)}<<a_{k+p\alpha\sigma_{1}}a^{+}_{p+q\nu\sigma_{2}}
a_{q\nu\sigma_{2}}|a^{+}_{k+s\beta\sigma_{3}}
a^{+}_{r\mu\sigma_{4}} a_{r+s\mu\sigma_{4}}>>^{(ir),p}) $$
For the explicit approximate calculation of the elements of the
self-energy it is convenient to write down the GFs in (54) in
terms of correlation functions by using the well-known spectral
theorem~\cite{tyab53}:
\begin{eqnarray}
(^{(ir)}<<a_{k+p\alpha\sigma_{1}}a^{+}_{p+q\nu\sigma_{2}}a_{q\nu\sigma_{2}}
|a^{+}_{k+s\beta\sigma_{3}}
a^{+}_{r\mu\sigma_{4}} a_{r+s\mu\sigma_{4}}>>^{(ir),p}) =
\nonumber\\
{1 \over 2\pi}\int_{-\infty}^{+\infty}{d\omega' \over \omega - \omega'}
(\exp(\beta \omega') +1)     \int_{-\infty}^{+\infty}\exp(-i\omega't)dt
\nonumber\\
<a^{+}_{k+s\beta\sigma_{3}}(t)
a^{+}_{r\mu\sigma_{4}}(t) a_{r+s\mu\sigma_{4}}(t)|
a_{k+p\alpha\sigma_{1}}a^{+}_{p+q\nu\sigma_{2}}a_{q\nu\sigma_{2}}
>^{(ir),p})
\end{eqnarray}
Further insight is gained if we select the suitable relevant ``trial"
approximation for the correlation function on the r.h.s. of (55). In
this paper we show that the earlier formulations, based on the
decoupling or/and on diagrammatic methods can be arrived at from our
technique but in a self- consistent way. Clearly the choice of the
relevant trial approximation for the correlation function in (55) can be
done in a few ways. For example, a reasonable and workable one may be
the following ``pair approximation" ~\cite{kuz3}, which is especially
suitable for low density of the quasiparticles:
\begin{eqnarray}
<a^{+}_{k+s\beta\sigma}(t)a^{+}_{r\mu-\sigma}(t)a_{r+s\mu-\sigma}(t)
a_{k+p\alpha\sigma}a^{+}_{p+q\nu-\sigma}a_{q\nu-\sigma}>^{ir} \approx
\nonumber\\
<a^{+}_{k+s\beta\sigma}(t)a_{k+p\alpha\sigma}>
<a^{+}_{r\mu-\sigma}(t)a_{q\nu-\sigma}>
<a_{r+s\mu-\sigma}(t)a^{+}_{p+q\nu-\sigma}>\nonumber\\
+
<a^{+}_{k+s\beta\sigma}(t)a_{q\nu-\sigma}>
<a^{+}_{r\mu-\sigma}(t)a_{k+p\alpha\sigma}>
<a_{r+s\mu-\sigma}(t)a^{+}_{p+q\nu-\sigma}> \end{eqnarray}
Using (56)
in (55) we obtain the approximate expression for the
self-energy operator in a self-consistent form (the self-consistency
means that we express approximately the self-energy operator in terms
of the initial GF and, in principle, one can obtain the required
solution by a suitable iteration procedure):  \begin{eqnarray}
M^{\alpha\beta}_{11}(k, \omega) = \frac{1}{N^2 \pi^3} \sum_{pq\mu\nu}
( B^{\alpha\nu}_{1}B^{\mu\beta}_{1} \int
\frac{d\omega_{1} d\omega_{2} d\omega_{3}}{\omega + \omega_{1} -
\omega_{2} - \omega_{3}} \nonumber\\
N(\omega_{1},\omega_{2},\omega_{3})
[g^{\mu\nu}_{p+q\uparrow\uparrow}(\omega_{1})
g^{\nu\mu}_{q\uparrow\uparrow}(\omega_{2})
g^{\alpha\beta}_{k+p\uparrow\uparrow}(\omega_{3}) +
g^{\alpha\mu}_{k+p\uparrow\uparrow}(\omega_{3})
g^{\nu\beta}_{q\uparrow\uparrow}(\omega_{2})
g^{\mu\nu}_{p+q\uparrow\uparrow}(\omega_{1})] +\nonumber\\
B^{\alpha\nu}_{1}B^{\mu\beta}_{2}
\int
\frac{d\omega_{1}d\omega_{2}d\omega_{3}}{\omega + \omega_{1} -
\omega_{2} - \omega_{3}}
~N(\omega_{1},\omega_{2},\omega_{3})
[(\downarrow\uparrow)(\uparrow\downarrow)(\uparrow\uparrow) +
(\uparrow\downarrow)(\downarrow\uparrow)(\uparrow\uparrow)]
+\nonumber\\
B^{\alpha\nu}_{2}B^{\mu\beta}_{1}
\int
\frac{d\omega_{1}d\omega_{2}d\omega_{3}}{\omega + \omega_{1} -
\omega_{2} - \omega_{3}}
~N(\omega_{1},\omega_{2},\omega_{3})
[(\uparrow\downarrow)(\downarrow\uparrow)(\uparrow\uparrow) +
(\uparrow\uparrow)(\uparrow\downarrow)(\downarrow\uparrow)]
+\nonumber\\
B^{\alpha\nu}_{2}B^{\mu\beta}_{2}
\int
\frac{d\omega_{1}d\omega_{2}d\omega_{3}}{\omega + \omega_{1} -
\omega_{2} - \omega_{3}}
~N(\omega_{1},\omega_{2},\omega_{3})
[(\downarrow\downarrow)(\downarrow\downarrow)(\uparrow\uparrow) +
(\uparrow\downarrow)(\downarrow\downarrow)(\downarrow\uparrow)])
\end{eqnarray}
where we have used the notations
$$N(\omega_{1},\omega_{2},\omega_{3}) = [n(\omega_{2})n(\omega_{3}) +
n(\omega_{1})(1 - n(\omega_{2}) - n(\omega_{3}))] ;$$
$$g_{k\sigma\sigma'}(\omega) = -{1 \over \pi}Im
G_{k\sigma\sigma'}(\omega + i\varepsilon); \quad n(\omega) =
[\exp(\beta\omega) + 1]^{-1}$$
Here we present for brevity the
explicit expression for a part of the Hamiltonian only without the last
term.
The equations (43) and (57) form a closed self-consistent system
of equations for the single-electron GF for the Hubbard model, but for
the weakly correlated limit only. In principle, one may use on the
r.h.s.  of (57) any workable first iteration-step forms of the GFs and
find a solution by repeated iterations.  It is most convenient to
choose as the first iteration step the following simple one-pole
approximation:  \begin{equation} g_{k\sigma}(\omega) \approx
\delta(\omega - \epsilon(k\sigma)) \end{equation} Then, using (58) in
(57), one can get an explicit expression for the
self-energy.  However, the actual explicit calculations will be much
more transparent if we confine ourselves to the single-orbital Hubbard
model in order to discuss more explicitly the reliability of the
present approach.  \section{Antiferromagnetic Single-Particle States}
The technique for obtaining of the antiferromagnetic solutions to the
correlated fermions on a lattice is presented in this section for the
single-orbital Hubbard model (10).  In general, it can be easily
applied for the multiorbital extended Hubbard model.\\ As discussed
above, the self-consistent approach to the calculation of the
one-particle properties requires the calculation of the folllowing GF
\begin{equation} \pmatrix{ <<a_{i\sigma}\vert a^{+}_{j\sigma}>> &
<<a_{i\sigma}\vert a^{+}_{j-\sigma}>> \cr <<a_{i-\sigma}\vert
a^{+}_{j\sigma}>> & <<a_{i-\sigma}\vert a^{+}_{j-\sigma}>> \cr} = \hat
G(ij;\omega) \end{equation} The equation of motion for the Fourier
transform of the GF has the form \begin{equation} \sum_{m} \hat A(im)
\hat G(mj;\omega) = \delta_{ij} \delta_{\sigma\sigma'} +  U
<<a_{i\sigma}n_{i-\sigma}|a^{+}_{j\sigma'}>>
\end{equation}
where
\begin{equation}
\hat A(im) =
\pmatrix{
(\omega\delta_{mi} -
t_{im})
& 0\cr
0&(\omega\delta_{mi} -
t_{im}) \cr}
\end{equation}
Using the definition of the irreducible parts (26)the equation of motion
can be exactly transformed to the following form
\begin{equation}
\sum_{m}
\hat A_{1}(im) \hat G(mj;\omega) = \delta_{ij}
\delta_{\sigma\sigma'} +  U \hat D^{ir}(ij;\omega)
\end{equation}
where
\begin{equation}
\hat A_{1}(im) =
\pmatrix{
(\omega\delta_{mi} -
t_{im} -U<n_{i-\sigma}>)
& -U<a_{i\sigma}a^{+}_{i-\sigma}>\cr
-U<a_{i-\sigma}a^{+}_{i\sigma}>&(\omega\delta_{mi} -
t_{im} - U<n_{i\sigma}>) \cr}
\end{equation}
To calculate the irreducible higher-order GF $D^{ir}$ we have to write
the equation of motion for it. After introducing the irreducible parts
for the operators in the right-hand-side we find
\begin{equation}
\sum_{n}
\hat D^{ir}(in;\omega)\hat A_{2}(nj) =  U^{2} \hat D_{1}(ij;\omega)
\end{equation}
where
\begin{equation}
\hat D_{1}(ij;\omega) =
\pmatrix{
^{(ir)}<<a_{i\sigma}n_{i-\sigma}|a^{+}_{j\sigma}n_{j-\sigma}>>^{(ir)}&
^{(ir)}<<a_{i\sigma}n_{i-\sigma}|a^{+}_{j-\sigma}n_{j\sigma}>>^{(ir)}\cr
^{(ir)}<<a_{i-\sigma}n_{i\sigma}|a^{+}_{j\sigma}n_{j-\sigma}>>^{(ir)}&
^{(ir)}<<a_{i-\sigma}n_{i\sigma}|a^{+}_{j-\sigma}n_{j\sigma}>>^{(ir)}\cr}
\end{equation}
Then equation of motion for the GF can be exactly transformed into the
following scattering equation
\begin{equation}
G(ij;\omega) = G_{0}(ij;\omega) +
\sum_{mn}G_{0}(im;\omega)
P(mn;\omega) G_{0}(nj;\omega)
\end{equation}
where the generalized mean-field GF $G_{0}$ reads
\begin{equation}
\sum_{m}A_{1}(im)G_{0}(mj;\omega) =
 \delta_{ij}
\end{equation}
and the scattering operator $P$ has the form
\begin{equation}
\hat P(ij;\omega) = U^{2}
\pmatrix{
^{(ir)}<<a_{i\sigma}n_{i-\sigma}|a^{+}_{j\sigma}n_{j-\sigma}>>^{(ir)}&
^{(ir)}<<a_{i\sigma}n_{i-\sigma}|a^{+}_{j-\sigma}n_{j\sigma}>>^{(ir)}\cr
^{(ir)}<<a_{i-\sigma}n_{i\sigma}|a^{+}_{j\sigma}n_{j-\sigma}>>^{(ir)}&
^{(ir)}<<a_{i-\sigma}n_{i\sigma}|a^{+}_{j-\sigma}n_{j\sigma}>>^{(ir)}\cr}
\end{equation}
The Dyson equation (41)  then will
be reduced for the single-band Hubbard model
to the following form \begin{equation} G(ij;\omega) =
G_{0}(ij;\omega) + \sum_{mn} G_{0}(im;\omega)M(mn;\omega)G(nj;\omega)
\end{equation}
The mass operator $M(mn;\omega) = U^{2}P^{(p)}(mn;\omega)$ describes
the inelastic (retarded) part of the electron-electron interaction. For
the purpose of analogy with the theory of
superconductivity~\cite{kuz19} let us write the Hartree-Fock (elastic)
part of the Coulomb mass operator (not included in (68)):
\begin{equation} \hat M^{HF}(im) = U \pmatrix{ <n_{i-\sigma}> &
<a_{i\sigma}a^{+}_{i-\sigma}>\cr <a_{i-\sigma}a^{+}_{i\sigma}>&
<n_{i\sigma}>) \cr}\delta_{im} \end{equation} To obtain workable
expressions for various parts of the mass operator we use the spectral
theorem, inverse Fourier transformation and make relevant approximation
in the time correlation functions. In analogy with the theory of
superconductivity the suitable approximation which describe the
interaction between the charge and spin collective excitations can be
written as \begin{eqnarray}
<a^{+}_{n\sigma}(t)a^{+}_{n-\sigma}(t)a_{n-\sigma}(t)
a_{m\sigma}a^{+}_{m-\sigma}a_{m-\sigma}>^{ir} \approx
\nonumber\\
<a^{+}_{n\sigma}(t)a_{m\sigma}>
<n_{n-\sigma}(t)n_{m-\sigma}>
\nonumber\\
+
<a^{+}_{n-\sigma}(t)a_{m-\sigma}>
<a^{+}_{n\sigma}(t)a_{n-\sigma}(t)
a_{m\sigma}a^{+}_{m-\sigma}>
\nonumber\\
+
<a_{n-\sigma}(t)a^{+}_{m-\sigma}>
<a^{+}_{n\sigma}(t)a^{+}_{n-\sigma}(t)
a_{m\sigma}a_{m-\sigma}>
\nonumber\\
+
<a^{+}_{n\sigma}(t)a_{m-\sigma}>
<a^{+}_{n-\sigma}(t)a_{n-\sigma}(t)
a_{m\sigma}a^{+}_{m-\sigma}>
\nonumber\\ +
<a^{+}_{n-\sigma}(t)a_{m\sigma}>
<a^{+}_{n\sigma}(t)a_{n-\sigma}(t)
a^{+}_{m-\sigma}a_{m-\sigma}>
\nonumber\\ +
<a_{m-\sigma}(t)a_{m\sigma}>
<a^{+}_{n\sigma}(t)a^{+}_{n-\sigma}(t)
a^{+}_{m-\sigma}a_{m-\sigma}>
\end{eqnarray}
The suitable or relevant approximations follow from the concrete
physical conditions of the problem under consideration. We consider
here for illustration the contributions from charge and spin collective
degrees of freedom. We get
\begin{eqnarray}
M(ij;\omega) =\frac {U^{2}}{2\pi^{2}} \int_{-\infty}^{+\infty}
d\omega_{1}d\omega_{2} \frac {ctg \frac {\beta\omega_{1}}{2} +
tg\frac {\beta\omega_{2}}{2}}{\omega - \omega_{1} - \omega_{2}} \qquad
\qquad \qquad  \qquad \qquad \qquad \qquad \\
\nonumber
\Bigl(
\pmatrix{
Im<<n_{i-\sigma}|n_{j-\sigma}>>_{\omega_{1}}
Im<<a_{i\sigma}|a^{+}_{j\sigma}>>_{\omega_{2}}&
Im<<n_{i-\sigma}|n_{j\sigma}>>_{\omega_{1}}
Im<<a_{i\sigma}|a^{+}_{j-\sigma}>>_{\omega_{2}}\cr
Im<<n_{i\sigma}|n_{j-\sigma}>>_{\omega_{1}}
Im<<a_{i-\sigma}|a^{+}_{j\sigma}>>_{\omega_{2}}&
Im<<n_{i\sigma}|n_{j\sigma}>>_{\omega_{1}}
Im<<a_{i-\sigma}|a^{+}_{j-\sigma}>>_{\omega_{2}}\cr} +\\
\nonumber
\pmatrix{
Im<<S^{-\sigma}_{i}|S_{j}^{\sigma}>>_{\omega_{1}}
Im<<a_{i-\sigma}|a^{+}_{j-\sigma}>>_{\omega_{2}}&
Im<<S_{i}^{-\sigma}|S_{j}^{-\sigma}>>_{\omega_{1}}
Im<<a_{i-\sigma}|a^{+}_{j\sigma}>>_{\omega_{2}}\cr
Im<<S_{i}^{\sigma}|S_{j}^{\sigma}>>_{\omega_{1}}
Im<<a_{i\sigma}|a^{+}_{j-\sigma}>>_{\omega_{2}}&
Im<<S_{i}^{\sigma}|S_{j}^{-\sigma}>>_{\omega_{1}}
Im<<a_{i\sigma}|a^{+}_{j\sigma}>>_{\omega_{2}}\cr}
\Bigr)
\nonumber
\end{eqnarray}
It shows that it is possible to do all calculations in the localized
Wannier basis as we did in deriving the equations for the strong
coupling superconductivity in transition metals~\cite{kuz19}. This has
great advantage for consideration of disordered transition metal
alloys.\\ As for the translationally invariant crystal with broken
symmetry the following special Fourier transform should be performed
for the generalized mean-field GF $ G_{0}(ij;\omega) $ (67)
\begin{eqnarray}
G^{11}_{0}(ij;\omega) = \sum_{k} \exp {[ik(R_{i} -
R_{j})]}G^{11}_{0}(k;\omega)\\
\nonumber
G^{12}_{0}(ij;\omega) = \sum_{k} \exp {[ikR_{i} - i(k+Q)R_{j}]}
G^{12}_{0}(k;\omega)\\
\nonumber
G^{21}_{0}(ij;\omega) = \sum_{k} \exp {[i(k+Q)R_{i} - ikR_{j}]}
G^{21}_{0}(k;\omega)\\
\nonumber
G^{22}_{0}(ij;\omega) = \sum_{k} \exp {[i(k+Q)(R_{i} -
R_{j})]}G^{22}_{0}(k;\omega)
\end{eqnarray} The result of this
transformation is then \begin{equation} G_{0}
=\pmatrix{G^{11}_{0}&G^{12}_{0}\cr G^{21}_{0}&G^{22}_{0} \cr} =
\frac{\pmatrix{ \omega - E^{HF}_{\downarrow}(k+Q) & \Delta_{\uparrow
\downarrow}(k)\cr \Delta_{\downarrow \uparrow}(k)&\omega -
E^{HF}_{\uparrow}(k) \cr}}{ (\omega - E^{MF}_{1}(k))(\omega -
E^{MF}_{2}(k))} \end{equation} where
\begin{eqnarray} E^{HF}_{\sigma} =
\epsilon (k) + U<n_{\sigma}> \qquad \qquad \qquad\\
\nonumber
\Delta_{\sigma -\sigma}(k) = U \sum{i} \exp(ikR_{i})<a_{i\sigma}
a^{+}_{i -\sigma}> \qquad \quad\\
\nonumber
E^{MF}_{1,2} = \Bigl (\frac
{E^{HF}_{\uparrow}(k) + E^{HF}_{\downarrow}(k+Q)}{2} \pm \sqrt { \Bigl(
\frac {E^{HF}_{\uparrow}(k) - E^{HF}_{\downarrow}(k+Q)}{2} \Bigr)^{2} +
\Delta_{\uparrow \downarrow}(k) \Delta_{\downarrow \uparrow}(k)}\quad
\Bigr)
\end{eqnarray}
It is evident that one can define the Overhauser's angle $\theta_{k}$
\begin{equation}
\cos^{2} \theta_{k} = \frac {
\Delta_{\uparrow \downarrow}(k) \Delta_{\downarrow \uparrow}(k)}
{(\omega - E^{HF}_{\uparrow}(k))^{2} +
\Delta_{\uparrow \downarrow}(k) \Delta_{\downarrow \uparrow}(k)}
\end{equation}
In Overhauser's notations $
\Delta_{\uparrow \downarrow}(k) = \Delta_{\downarrow \uparrow}(k) =
\Delta $ . The self-consistent set of equations for determining of the
SDW (or "gap") order parameter $\Delta$ , chemical potential $\mu$  and
averaged moment $<s^{z}>$ is \begin{eqnarray}
\Delta = U/N \sum_{k}<a^{+}_{k+Q \downarrow}a_{k \uparrow}>\\
\nonumber
<s^{z}> = U/N \sum_{k}<a^{+}_{k \uparrow}a_{k \uparrow} -
a^{+}_{k \downarrow}a_{k \downarrow}>\\
\nonumber
n = N^{-1} \sum_{k}\Bigl( n(E^{MF}_{1}(k) + n(E^{MF}_{2}(k) \Bigr)
\nonumber
\end{eqnarray}
The above expressions were derived for correlated itinerant
fermions on a lattice within the Hubbard model and for finite
temperatures. These equations were also deduced in previous papers
in the course of their analysis. Here we deduced it by using more
sophisticated arguments of the IGFs method in complete analogy
with our description of the Heisenberg  antiferromagnet at finite
temperatures~\cite{kuz52}. However, the self-consistent system of
equations (69), (72) for determining the quasiparticle spectra
with damping is not as obvious as generalization as the equations
(77).  This is intrinsically the many-body manifestation of the
correlation effects of itinerant fermions on a lattice and shows
clearly the advantage of the present approach.\\ To confirm this,
the explicit calculation of the damping should be performed. The
natural way to tackle this program would then to look at the
calculations of the collective GFs of the generalized spin (and
charge) susceptibilities in (72) but this deserves of separate
consideration. Again this problem bears close similarity to the
paramagnetic Hubbard model and the antiferromagnetic Heisenberg
model and it can be argued that this effect of interference of
single-particle and collective modes of excitations should be
considered carefully.  \section{Discussion} We have been concerned
in this paper with establishing  the essence of single-particle
excitations of correlated lattice fermions, rather than with their
detailed properties. We have considered the single- and multiband
Hubbard model but the calculational details were mainly presented
for the single-band Hubbard model where the appropriate concepts
are easier to demonstrate.  We have considered a general family of
symmetry broken solutions for itinerant lattice fermions,
identifying the type of ordered states and then derived explicitly
the functional of generalized mean fields and the self-consistent
set of equations which describe the quasiparticle spectra and
their damping in the most general way. While such generality is
not so obvious in all applications, it is highly desirable in
treatments of such complicated problems as the competition and
interplay of antiferromagnetism and superconductivity, heavy
fermions and antiferromagnetism etc., because of the non-trivial
character of coupled equations which occur there. Both these
problems are subject of current but independent research.\\
Another development of the present approach is the consideration
of the itinerant antiferromagnetism of highly correlated lattice
fermions when $U$ is very large but finite. Like the
weakly-coupled case described in this paper, the symmetry broken
approach will work, but matters are complicated by the necessity
of constructing  the more extended algebra of relevant
operators~\cite{kuz3}. This idea has been carried out for the
paramagnetic solution of the single-band Hubbard
model~\cite{kuz10}. It would be interesting to understand on a
deeper level the relationship between Mott-Hubbard metal-insulator
transition and various ordered magnetic states within the Hubbard
model.\\ In conclusion, we have demonstrated that the Irreducible
Green's Functions approach is a workable and efficient scheme for
the consistent description of the correlated fermions on a lattice
at finite temperatures and that it can be generalized naturally to
include the symmetry broken concept.
%

%

\begin{thebibliography} {99}
%
\bibitem{kuz1} A. L. Kuzemsky, in:
Superconductivity and Strongly Correlated Electron Systems, eds.
C. Noce et al (World Scientific, Singapoure, 1994) p.346.
\bibitem{bick2} N. E. Bickers and D. J. Scalapino, Ann.Phys.(N.Y.)
{\bf 193} (1989) 206
\bibitem{kuz3}
A. L. Kuzemsky, Molecular Phys.Rep. {\bf 17} (1997) 221
\bibitem{geb4}
F. Gebhard, The Mott Metal-Insulator Transition, Springer Tracts
in Modern Physics {\bf 137}, (Springer Verlag, Berlin, 1997).
\bibitem{ruv5} J. Ruvalds,
Supercond.Sci.Technol. {\bf 9} (1996) 905
\bibitem{pin6} A. Sokol and D. Pines, Phys.Rev.Lett.
{\bf 71} (1993) 2813
\bibitem{kampf7} A. Kampf, Phys.Rep. {\bf 249}
(1994) 219
\bibitem{kuz8} A. L. Kuzemsky, Phys.Lett. {\bf A 153} (1991)
466.
\bibitem{kuz9} A. L. Kuzemsky, J.-C. Parlebas and H. Beck, Physica
{\bf A 198} (1993) 606.
\bibitem{kuz10} A. L. Kuzemsky, Nuovo Cimento
{\bf B 109} (1994) 829.
\bibitem{kuz11} A. L. Kuzemsky, Intern.J.Modern
Phys. {\bf B 10} (1996) 1895
\bibitem{kem12} G. Kemeny, Ann.Phys.(N.Y.)
{\bf 32} (1965) 404
\bibitem{joh13} B. Johansson, Ann.Phys.(N.Y.) {\bf
74} (1972) 369
\bibitem{lieb14} V. Bach, E. H. Lieb and J. P. Solovej,
J.Stat.Phys. {\bf 76} (1994) 3.
\bibitem{fer15} J. Ferrer, Phys.Rev.
{\bf B 51} (1995) 8310.
\bibitem{alt16} J. Altmann, W. Brenig, A. P. Kampf
and E. Muller-Hartmann, Phys.Rev. {\bf B 52} (1995) 7395.
\bibitem{kampf17}
A. P. Kampf, Phys.Rev. {\bf B53} (1996) 747.
\bibitem{kuz18}
A. L. Kuzemsky, Doklady Acad.Nauk SSSR {\bf 309} (1989) 323.
\bibitem{kuz19}
G. Vujicic, A. L. Kuzemsky and N. M. Plakida, Theor.Math.Phys.
{\bf 53} (1982) 138
\bibitem{kuz20} A. L. Kuzemsky, A. Holas and N. M. Plakida,
Physica {\bf B 122} (1983) 168
\bibitem{raja21} A. K. Rajagopal,
Phys.Rev.  {\bf 142} (1966) 152
\bibitem{her22} C. Herring, in:
"Magnetism" (G. Rado and H. Suhl Eds.), Academic Press, N.Y. 1966,
vol.IV
\bibitem{slat23} J. C. Slater, Phys.Rev. {\bf 82} (1951) 538.
\bibitem{declo24} J. des Cloizeaux, J.de Physique {\bf 20} (1959)
606; 751.
\bibitem{over25} A. W. Overhauser, Phys.Rev.Lett. {\bf 3} (1959) 414.
\bibitem{over26} A. W. Overhauser, Phys.Rev.Lett. {\bf 4} (1960) 462.
\bibitem{vig27} O. Fajen and G. Vignale, Solid State Commun. {\bf 77}
(1991) 829.
\bibitem{over28} A. W. Overhauser, Phys.Rev. {\bf 128} (1962) 1347.
\bibitem{ham29}
D.R.Hamann and A. W.  Overhauser, Phys.Rev. {\bf 143} (1966) 183.
\bibitem{bayl30}
M. Baylin, Phys.Rev. {\bf 136} (1964) 1321;\\A. W. Overhauser,
Phys.Rev. {\bf B 10} (1974) 4918.
\bibitem{berg31}
K. F. Bergren and B. Johansson, Physica {\bf 40} (1968) 277
\bibitem{cade32} N. A. Cade and W. Young,
Adv.Phys.  {\bf 26} (1977) 393.
\bibitem{faw33}
E. Fawcett, Rev.Mod.Phys. {\bf 60} (1988) 209
\bibitem{fed34} P.A.Fedders and
P. C. Martin, Phys.Rev. {\bf 143} (1966) 245.
\bibitem{call35}
J. Callaway and D. G. Kanhere, Phys.Rev. {\bf B 49} (1994) 12823
\bibitem{hub36} J.Hubbard,
Proc.Roy.Soc. {\bf A 276} (1963) 238.
\bibitem{penn37} D. R. Penn,
Phys.Rev.  {\bf 142} (1966) 350.
\bibitem{adam38}
L. Adamowicz, R. Swirkowicz ans A. Sukiennicki, Physica {\bf 75}
(1974) 613
\bibitem{nolt39}
W. Nolting, Physica {\bf 80B} (1975) 303
\bibitem{oles40} A. M. Oles, Phys.Rev. {\bf B 28} (1983) 327.
\bibitem{kris41}
O. Krisement, Z.Physik {\bf 270} (1974) 383.
\bibitem{bog42} N. N.  Bogolubov, Physica
{\bf 26} (1960) S1.
\bibitem{ster43} H. Stern and R. Brout, Physica {\bf 30} (1964)
1689.
\bibitem{ster44} H. Stern, Phys.Rev. {\bf 147} (1966) 94.
\bibitem{schn45} T. Schneider and P. F. Meier, Physica {\bf 67} (1973)
521.
\bibitem{tas46} H. Tasaki, J.Stat.Phys. {\bf 76} (1994) 745.
\bibitem{mat47}
R.D.Mattuck and B. Johansson, Adv.Phys. {\bf 17} (1968) 509.
\bibitem{tas48}
H. Tasaki, J.Stat.Phys. {\bf 84} (1996) 535.
\bibitem{liu49}
R. S. Fishman and S. H. Liu, Phys.Rev.Lett.{\bf 76} (1996) 2398.
\bibitem{kis50}
R. Kishore and S. K. Joshi, J.Phys.C:Solid State Phys. {\bf4}
(1971) 2475.
\bibitem{oza51}
S. Yamamoto and M. Ozaki, Solid State Commun. {\bf 83} (1992)
329;335
\bibitem{kuz52}
A. L. Kuzemsky and D. Marvakov, Theor.Math.Phys. {\bf 83} (1990)
147
\bibitem{tyab53}
S. V. Tyablicov, "Methods in the Quantum Theory of Magnetism"
Plenum Press, New York, 1967.
%
\end{thebibliography}
\end{document}